\documentstyle[12pt,epsf]{article}
\newcommand{\be}{\begin{equation}}
\newcommand{\ee}{\end{equation}}
\newcommand{\bea}{\begin{eqnarray}}
\newcommand{\eea}{\end{eqnarray}}
\newcommand{\ci}{\cite}
\newcommand{\bi}{\bibitem}
\newcommand{\nono}{\nonumber \\}

\def\dal{\,\lower0.3ex\vbox{\hrule\hbox{\vrule\kern2pt\vbox{\kern4pt\kern4pt}
\kern2pt\vrule}\hrule}\,}

\begin{document}

\title{\sl Coherent polychotomous waves from an attractive well}
\vspace{1 true cm}
\author{G. K\"ALBERMANN\thanks{Permanent address: Faculty of
Agriculture and Racah Institute of Physics, Hebrew University, Jerusalem 91904
, Israel}\\
Cyclotron Institute, 
Texas A\&M University\\
College Station, TX 77843, USA\\}

\maketitle

\begin{abstract}
\baselineskip 1.5 pc
A novel effect of a wave packet scattering off an attractive one-
dimensional well is found numerically and analytically.
For a wave packet narrower than the width of the well, the
scattering proceeds through a quasi-bound state of almost zero energy.
The wave reflected from the well is a polychotomous (multiple peak)
monochromatic and coherent train. 
The transmitted wave is a spreading smooth wave packet.
The effect is strong for low average speeds of the packet,
and it disappears for wide packets.
\end{abstract}

\newpage
\baselineskip 1.5 pc

This paper deals with the classical textbook exercise of a wave packet
interacting with an attractive well\ci{merz}.
Despite being a thoroughly studied example of quantum
scattering for plane wave stationary states, 
the effect to be presented here for wave packets
was yet to be found.

A one-dimensional attractive well can either reflect or transmit a
wave.  Reflection and transmission coefficients are the simplest
scattering amplitudes. They can easily be calculated for a square
well by using plane waves and elementary continuity conditions.
The analysis of the exact time development of a packet, as well as the
treatment of realistic well shapes, is however reserved for numerical treatment.

We here show that wave packet scattering possesses an intriguing
aspect: Packets that are narrower than the well width initially,
resonate inside it, generating a reflected wave that is
coherent and monochromatic in amplitude, a polychotomous wave train.

Polychotomous (multipeak) waves are observed
when a superintense laser field focuses on an atom\ci{grobe}.
Ionization is hindered and the wave function is localized, in spite of
the presence of 
the strong radiation field. The wave packet representing
the excited electron eventually spreads and the degree of
localization and/or ionization depends on the parameters of
the radiation field. 
The above effect appears when the external field operates
on a bound state.

We will describe here a similar and quite unexpected phenomenon when
a wave packet scatters off an attractive well. 
Localization of the reflected waves will be also found.
These waves spend a large amount of time spreading out of the scattering
region.
The speed of the reflected wave is independent of the initial
average energy of the packet. It is completely
generated by a bound state of almost zero binding energy.
The well acts as a resonator that emits a coherent wave, but, only
backwards.

The effect may be tested in back angle nuclear reactions and
an estimate of the energy, projectiles and candidate targets will
be given below. The effect is analogous to lasing inside a
cavity, the well becomes then the most natural laser available.

Consider a minimal uncertainty wave packet traveling from the left
with an average speed $v$, initial location $x_0$, mass $m$, wave number
$q~=~m~v$ and initial width $\delta$, 

\bea\label{packet}
\psi~=~C~exp\bigg({i~q~(x-x_0)-\frac{(x-x_0)^2}{4~\delta^2}}\bigg)
\eea

impinging on an attractive well located at the origin, with depth $A$ and width
$w$. For the sake of simplicity we use an exponential well, but
the results are not specific to the type or shape of the well

\bea\label{well}
V(x)=-A~exp\bigg({-\frac{x^2}{w^2}}\bigg)
\eea

We solve the Schr\"odinger equation for the scattering
event in coordinate space taking care of unitarity.
We use the method of Goldberg et al.\ci{gold}, that proved to
be extremely robust and conserves the wave normalization with an error of less
than 0.01 \%, even after hundreds of thousands of time step iterations.
We have verified that the solutions actually solve the equation with
extreme accuracy by explicit substitution.
Other simple discretization methods of resolution such as Runge-Kutta, leapfrog,
etc., are unstable for this type of equation, they violate unitarity.

We study the scattering of an impinging packet with $\delta=0.5$ and
a well width of $w=1$. We also use a large mass $m=20$ in order
to prevent the packet from spreading too fast \ci{merz}.
\begin{figure}[tb]
\epsffile{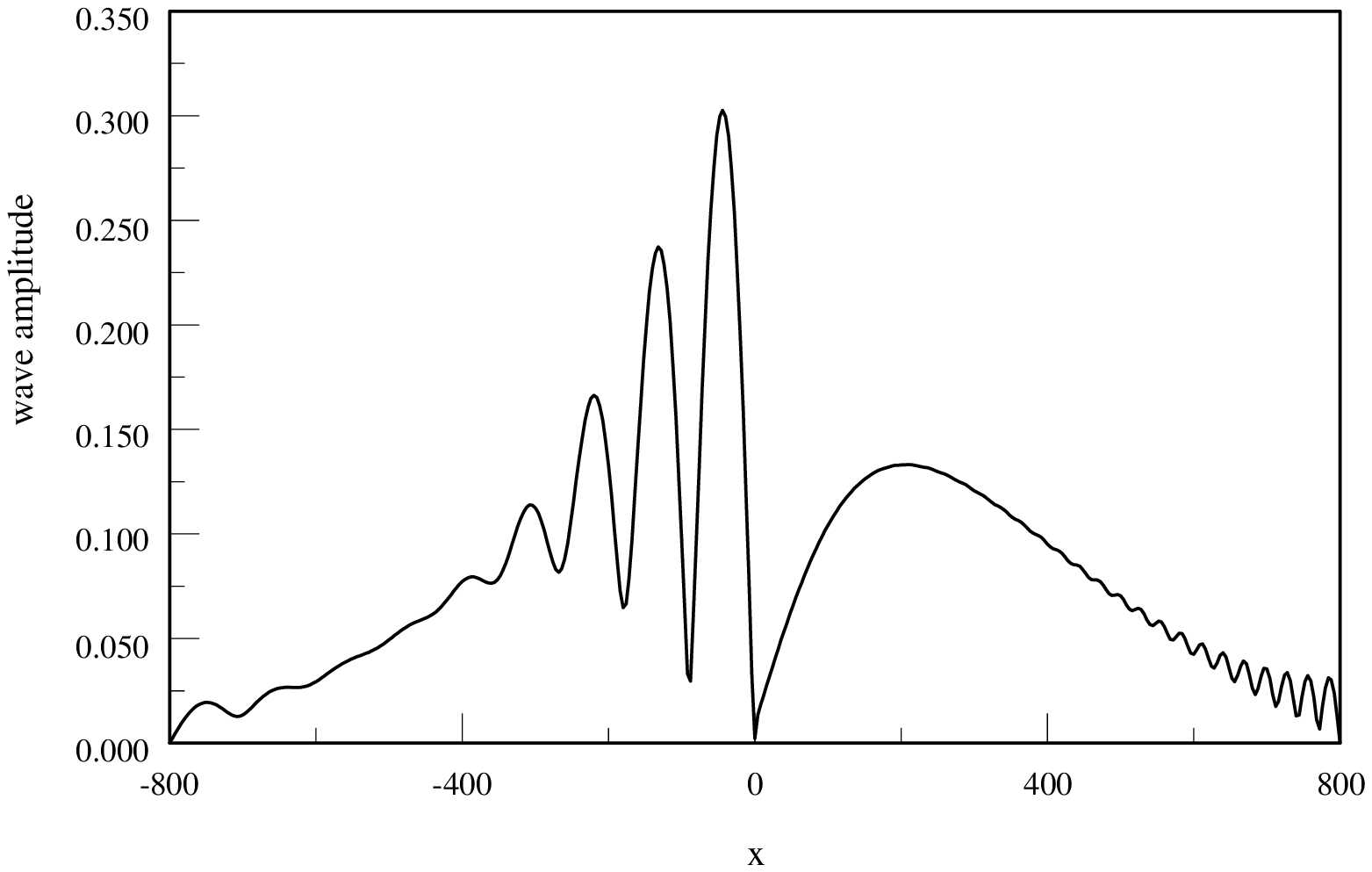}
\vsize=5 cm
\caption{\sl Wave amplitude as a function distance x for an initial 
wave packet of width $\delta=0.5$ starting at $x_0$=-10 impinging upon a well 
of width, w=1 and depth A=1 after t=5000, the initial
average momentum of the packet is q=0.2}
\label{fig1}
\end{figure}
\begin{figure}[tb]
\epsffile{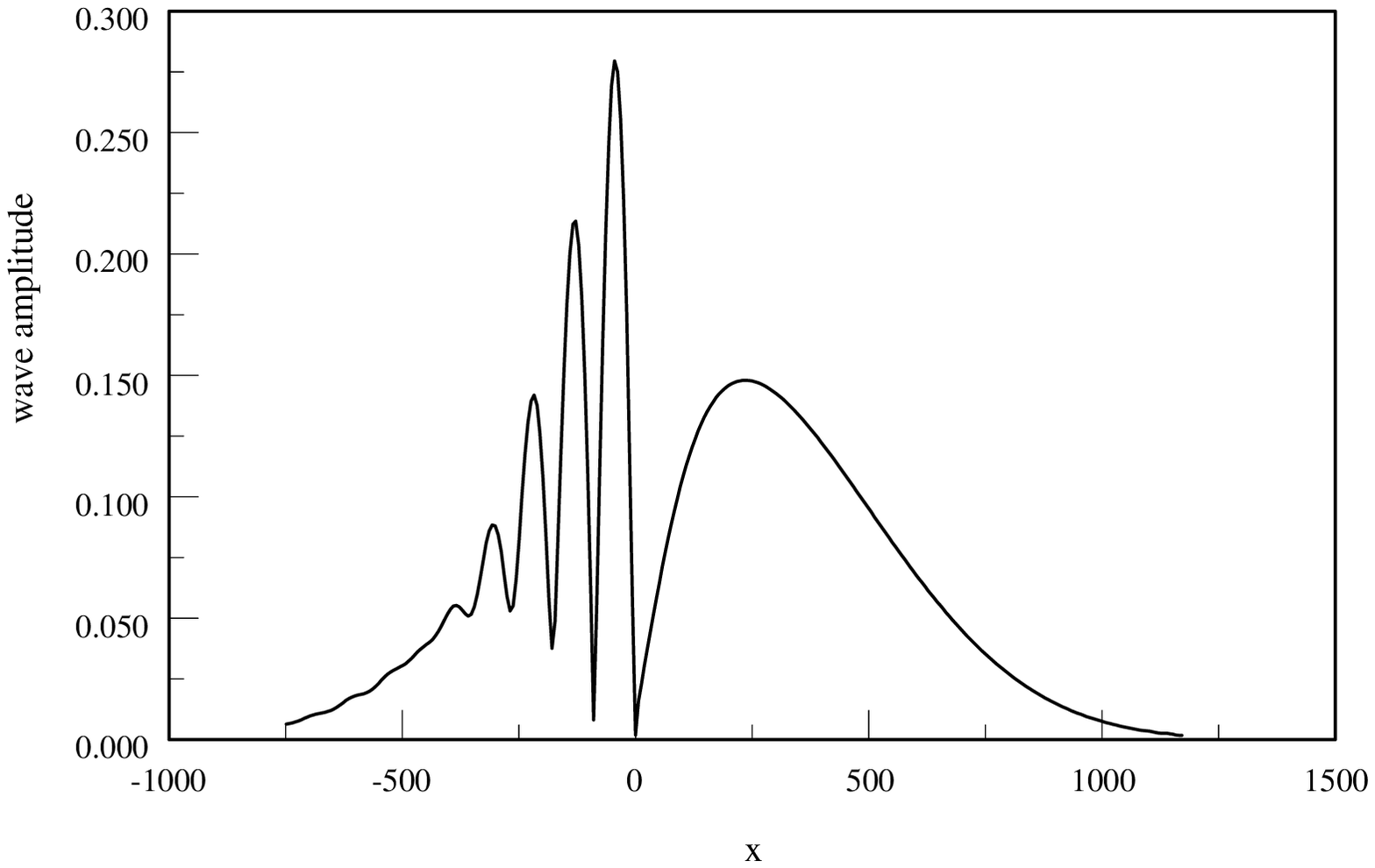}
\vsize=5 cm
\caption{\sl Same as figure 1 but for an initial momentum q=0.6}
\label{fig2}
\end{figure}
\begin{figure}[tb]
\epsffile{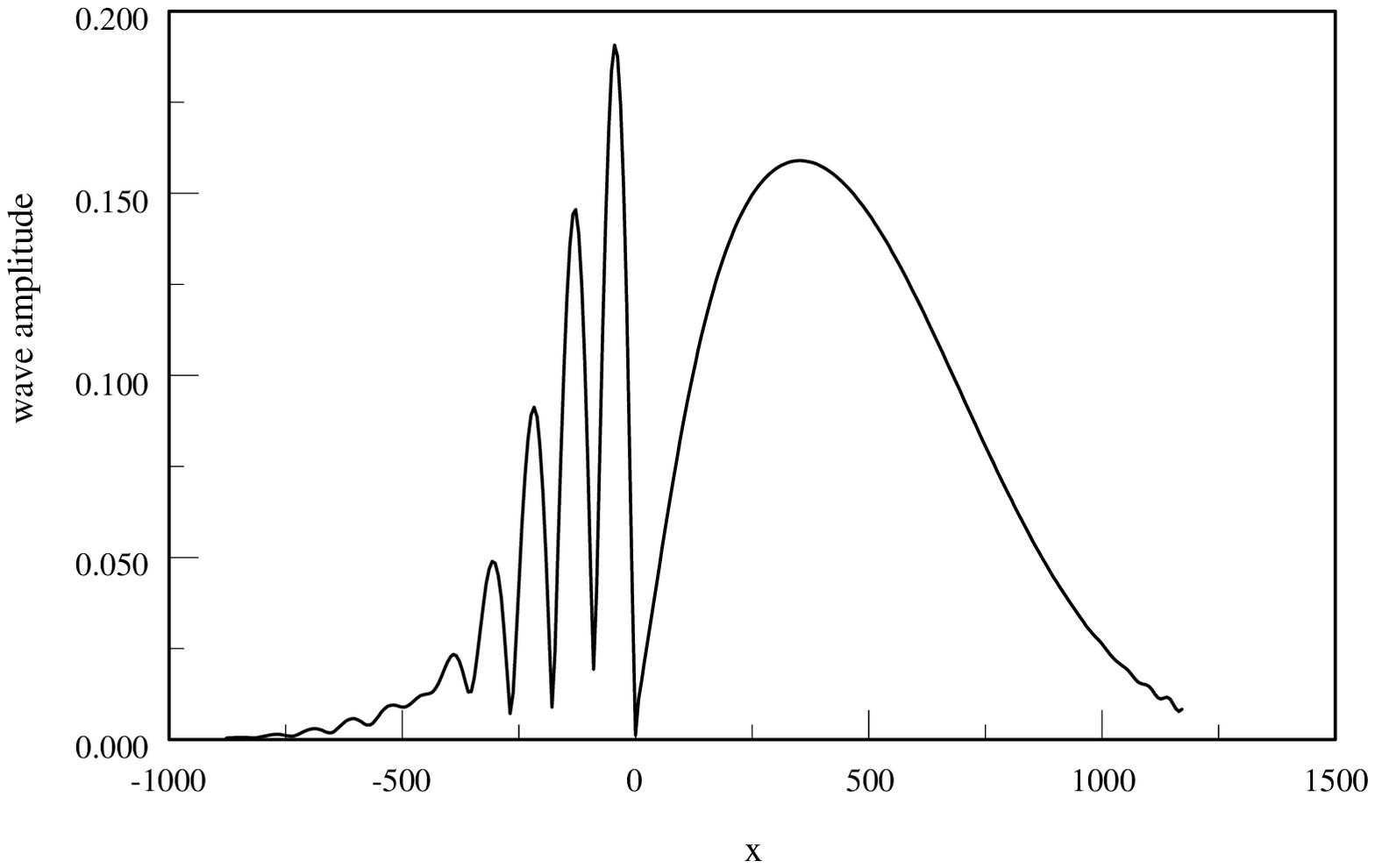}
\vsize=5 cm
\caption{\sl Same as figure 1 but for an initial momentum q=1.4}
\label{fig3}
\end{figure}
\begin{figure}[tb]
\epsffile{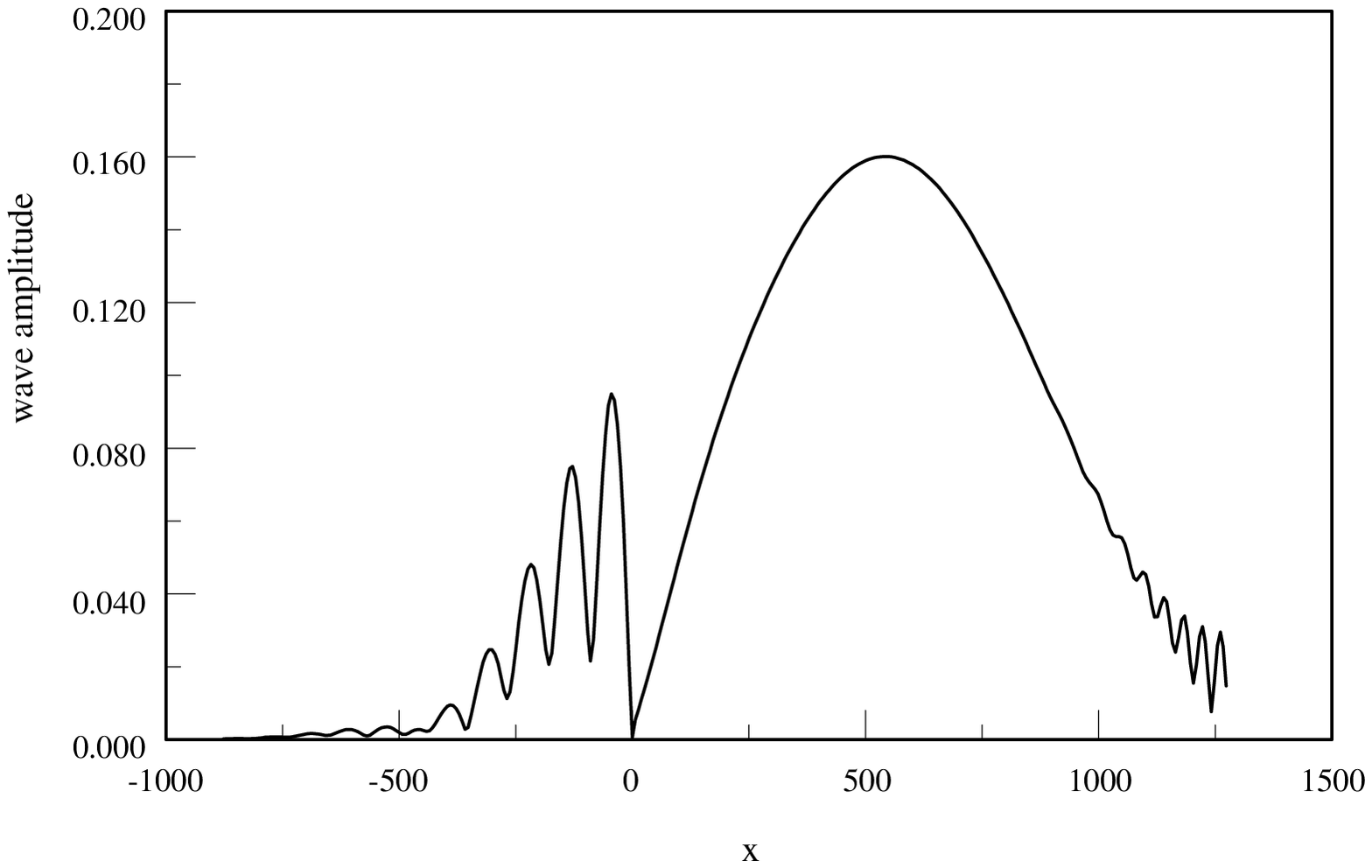}
\vsize=5 cm
\caption{\sl Same as figure 1 but for an initial momentum q=2.2}
\label{fig4}
\end{figure}
Figures 1 through 4  show pictures of the reflected and transmitted waves
after long periods of time for various initial velocities. The ripples at the
left and right ends of the figures are due to reflection from the boundaries.

We would have expected a smooth
packet for each the reflected and transmitted waves.
This is not the case in the figures.
A polychotomous (multiple peak) wave recedes from the well.
For low velocities, corresponding
to average packet energies less than half the well depth, 
several peaks in the reflected wave show up. 
Simple inspection reveals that the distance
between the peaks is constant. 
The reflected wave is propagating with an amplitude of the form

\bea\label{amplitude}
C(x)\approx e^{-\lambda |x|}~sin^2(kx)
\eea

The exponential drop is characteristic of a bound state solution inside the 
well.
The parameters $\lambda$ ad $\sl k$, are independent of the initial velocity, 
but depend on time. The wave spreads and its amplitude diminishes, as expected.
We have checked that the polychotomous behavior continues for
$t\rightarrow\infty$ without modification.

The only possible explanation for both the coherence and
independence of the reflected wave on the initial energy, is
that it must be generated by a resonant phenomenon inside the well.
\begin{figure}[tb]
\epsffile{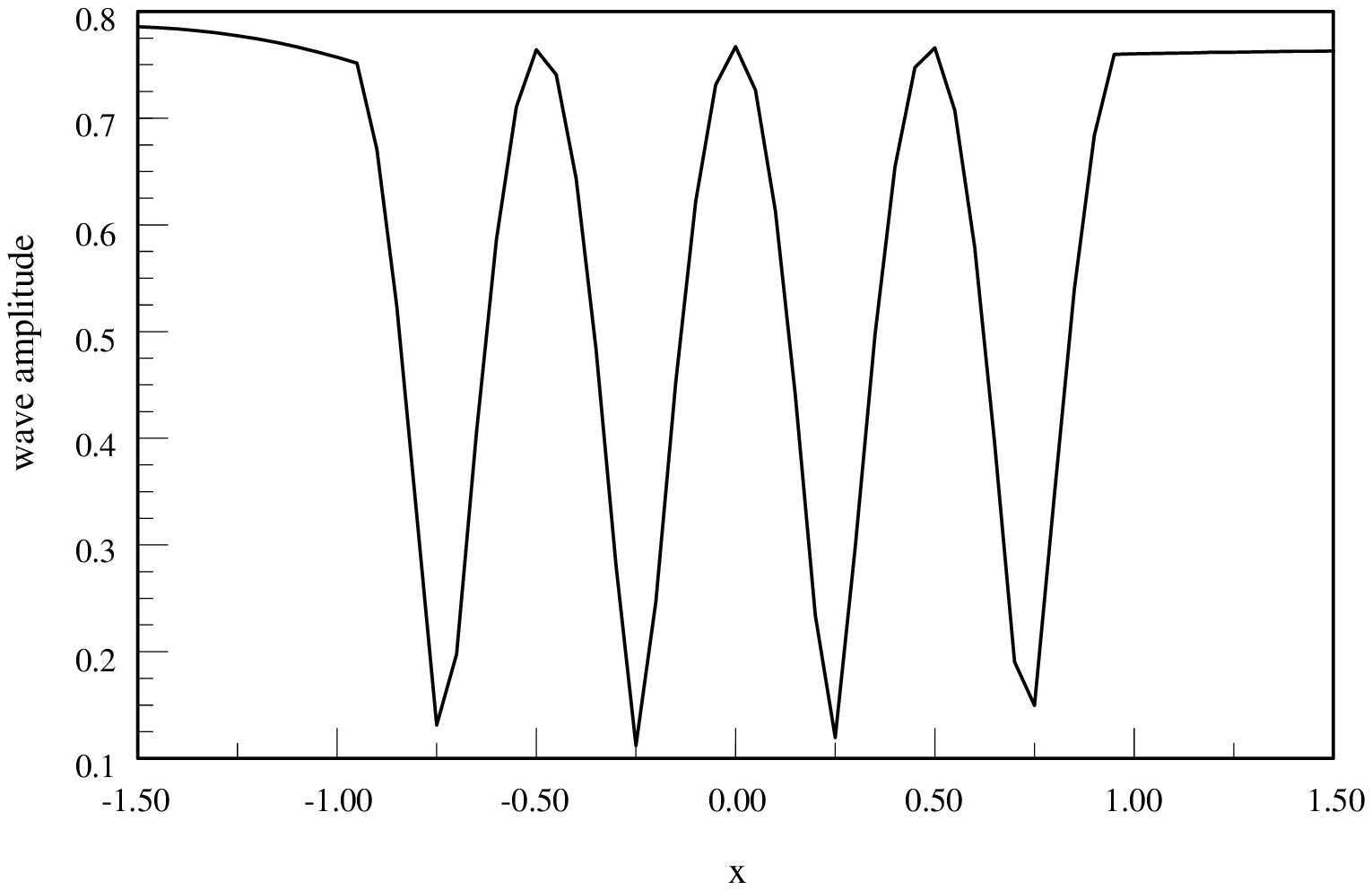}
\vsize=5 cm
\caption{\sl Same as figure 1, for an initial momentum q=1 showing the 
resonance inside the well and for t=200, at which the resonance is fully
developed}
\label{fig5}
\end{figure}
\begin{figure}[tb]
\epsffile{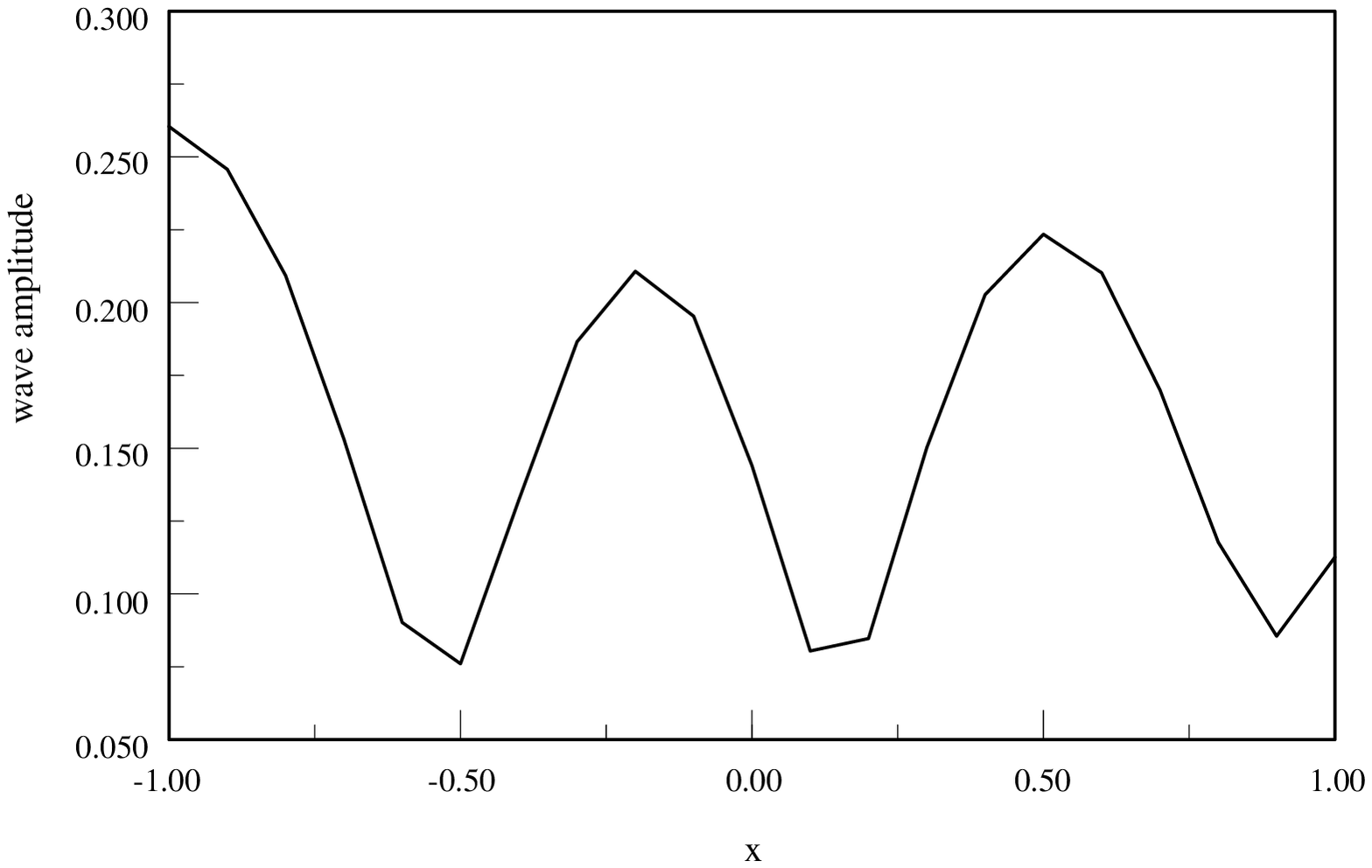}
\vsize=5 cm
\caption{\sl Same as figure 5,but for a mass m=11}
\label{fig6}
\end{figure}
Figure 5 shows the excitation of the resonance inside the well for
an initial velocity of $v=0.05$.
The same picture arises for the different velocities.
The resonance is a quasi-bound state. The energy of the bound state
is the closest it can be to zero.
If we borrow the bound state conditions from the case of a square well
for even and odd states

\bea\label{condition}
k'tan(k'w)=~k, for~even~states\nono
k'cot(k'w)=~-k, for~odd~states
\eea

where $k'=~\sqrt{2~m~(A+E)}$,$\sl E$ is the energy of the bound state, 
$\sl A$ the depth of the well, and $k=~\sqrt{2~m~|E|}$; we find 
for $k\approx~0$,   $k'w=n\pi$ for even states, and $k'w=(2n+1)\pi/2$
for odd states. 

For the parameters used here  $k'\approx 2\pi$.
This is the wavenumber that can be read off by
inspecting figure 5. (Note that the figure displays the absolute
value of the wave). 
The bound state condition is indeed operative. We may further
put this in evidence, by modifying the mass of the packet to $m\approx 11$.
For this mass, the bound state condition is satisfied only with
an odd state for which $k'w\approx \frac{3\pi}{2}$. 
Figure 6 reveals that the condition is again correct.

The state excited inside the well is a quasi-bound state
that decreases in amplitude as a function of time (decays) emitting energy to
both sides of the well. However, its shape remains constant
even at extremely large times.
In a particulate point of view,
the resonance in the well produces a coherent bunch of particles.
Both the particulate aspect and the wave aspect through
the modulation of the wave amplitude are still present.

The real and imaginary parts of the packet inside the well 
follow the trend of the amplitude depicted in the figures, their separate
amplitudes do not always coincide, and their phases may differ as 
time passes. These differences may be traced back to the nonresonant components
of the scattering that show up as a background under the peaks of the
reflected waves. This nonresonant background is generated by the coupling to
the infinite tower of free modes in the well and to the other bound states
besides the dominant one at zero energy.

From figures 1-4, it is clear that the reflected coherent wave becomes
less and less noticeable as we increase the initial energy of the 
wave packet. For average packet energies of the order of $\sl A$, the well 
depth, the reflected wave
is negligible, but, it still shows a tiny coherent train.
The fact that the wavenumber of the reflected wave is independent
of speed (as well as packet amplitude) suggests that it is also
constrained by the resonance condition. Trial and error lead us to
a simple formula.
We find that the condition for the reflected wave wavenumber at the time of 
formation (later it will decrease due to packet spreading) is
$ 4~k~w=\pi$.

Resorting to classical wave phenomena we could argue
that the reflected wave has a virtual piece inside the well. 
Constructive interference between this virtual wave reflected at the
right boundary with a phase shift of $\pi$, 
with the wave reflected from the left boundary
of the well is exactly the equation above.

The transmitted packet travels at a different velocity than that the coherent
reflected packet. It may be easily read out from the graphs following
the development in time, or, by taking a large enough time for which the well
is far behind.
The reflected packet travels with a constant
speed of $v=k(t_{formation})/m$. 
The speed can be found by evaluating the effective
center of mass {\bf X} 

\bea\label{cm}
X_{refl}=\int_{-\infty}^{-w}{dx~x~|\psi(x)|^{2}}
\eea
Where the wave function is properly normalized to 1.
Using the above equation one finds that the reflected wave center of mass
recedes with a constant speed of approximately $v=0.03$ independent of
the initial speed of the incident packet, while the transmitted wave
rides away with a velocity slightly higher than the initial packet average
velocity, and it is determined by overall energy conservation.

The polychotomous effect disappears when the wave packet is broader than 
the well.
Figure 7 shows the standard reflected and transmitted waves
as usually seen in the literature\ci{gold} for a wave packet of $\delta=
2$ much greater than the well width of $w=1$.
No coherent reflected wave is seen.

The existence or inexistence of the polychotomous wave
depends only on the $\sl initial$ packet width. This is somewhat surprising
because for very low speeds, the time taken by the packet to reach the well,
if considered as a classical object, would be so large and the spreading
so important that its extent would exceed by far the well width. Nevertheless,
this seems to be an unimportant fact. The only thing that matters is the
initial spread of the packet. Following the evolution of the scattering
in short time lapses, one notices right away that the reflected and transmitted
waves are created almost immediately after the process starts.
The well senses the packet width from far away.
\begin{figure}[tb]
\epsffile{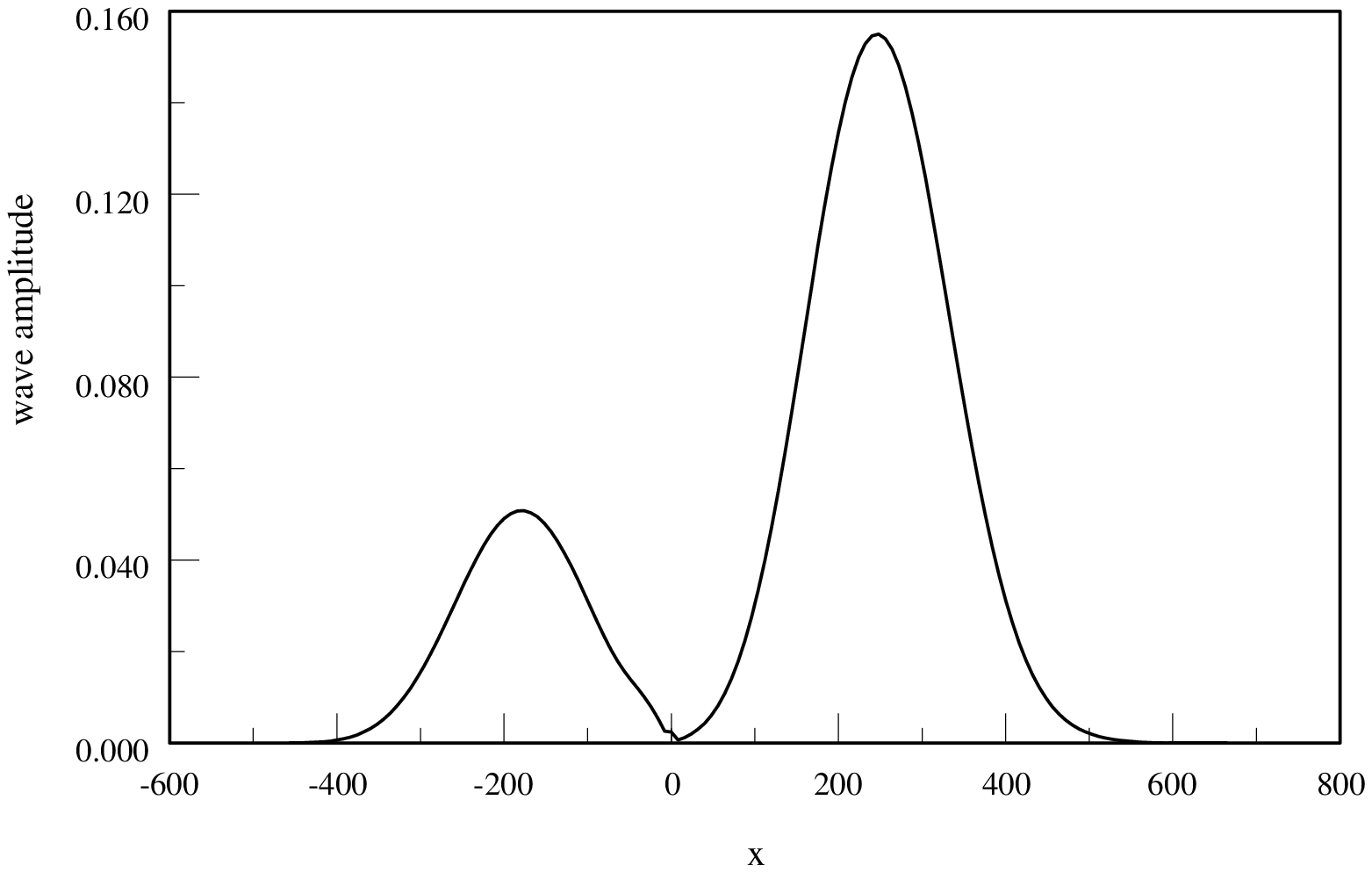}
\vsize=5 cm
\caption{\sl Same as figure 1 but a wave packet of width $\delta=2$ and q=1}
\label{fig7}
\end{figure}

We have also investigated other types of wells, such as a
Lorentzian well, a square well, etc.,
and found the same phenomena described here.
Moreover the effect is independent of the shape of the packet
as long as it is narrower than the well width.
We have used square packets, Lorentzian
packets, linear exponential packets, etc., with analogous results.

In order to check the effect analytically we resort to the
simplest packet, namely a square packet 

\bea\label{square}
\psi(x)=e^{i~q~(x-x_0)}~\Theta(d-|x-x_0|)
\eea

where $d$ is half the width of the packet, $x_0$ the initial position
and $q$ the wave number.
The wave impinges on a square well located at the origin, whose
width is $2a$ and depth $V_0$

\bea\label{sqwell}
V(x)= -V_0~\Theta(a-|x|)
\eea

This case is solvable using the techniques of ref.\ci{zidell}.
The method is appropriate only for packets with sharp edges, that
terminate at a certain point. It consists of integrating the Fourier amplitude
of the wave using a contour in the complex momentum plane
that avoids the poles of the scattering matrix corresponding to the
bound states. For each momentum, one uses the appropriate stationary
scattering state for the square well.
The integral reads
\bea\label{integral}
\psi(x,t)=~\int_{\sl C}\phi(x,p)~a(p,q)~dp
\eea
where $\sl C$ is a contour that goes from $-\infty$ to $+\infty$
and circumvents the poles that are on the imaginary axis 
for $p~<~i\sqrt{2 m~V_o}$ by closing it above them.
$\phi(x,p)$ is the stationary solution to the square well scattering
problem for each $\sl p$ and $\sl a(p,q)$ is the Fourier transform
amplitude for the initial wave function with average momentum $\sl q$.
Figure 8 shows the results of eq.~(\ref{integral}) for the
reflected and transmitted waves for a scattering
of an initial wave packet with average momentum $q=1$ 
packet width $d=0.5$, and well parameters $V_0=1,~a=1$ initially
far away from the well at $x_0=~-10$ (an essential condition
for the integral to converge). The reflected wave shows exactly the same 
polychotomous behavior as the numerical simulations.
The effect is general, even the packet amplitude is unimportant.
\begin{figure}[tb]
\epsffile{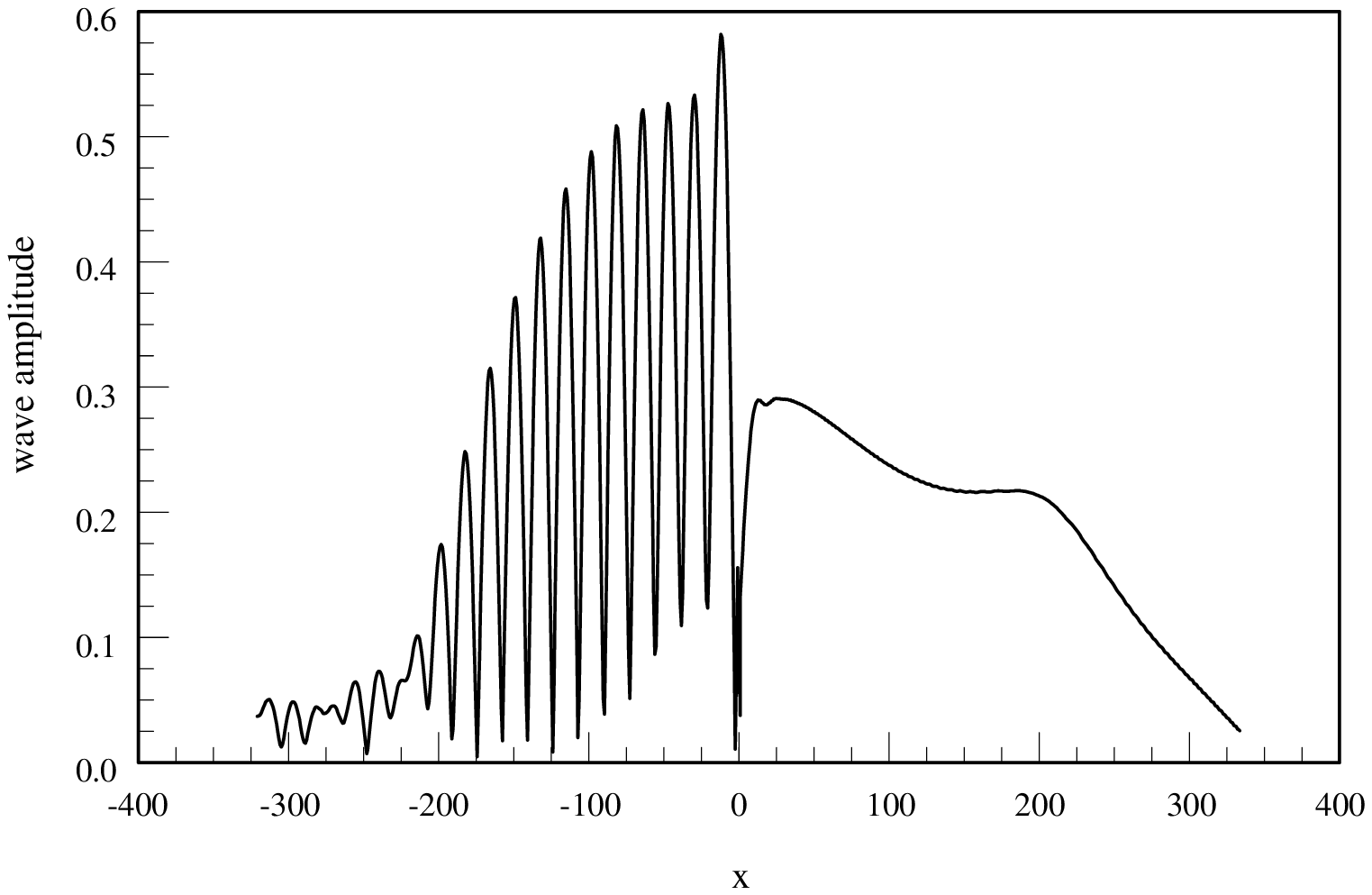}
\vsize=5 cm
\caption{\sl Theoretical calculation for a square initial packet
scattering off a square well for t=1000}
\label{fig8}
\end{figure}

Let us consider now the conditions for the effect to be measured
experimentally.
Consider for example backward angle scattering of
neutrons, or protons on nuclei.
Although our treatment was one-dimensional, it should apply also
for the case of zero angular momentum in three dimensions.
Typical nuclear well depths are around 30 MeV, with widths
of around a few Fermi for light nuclei, hence $k'\approx 1.25 fm^{-1}$.
The condition for the excitation of
the metastable resonance in the well and the coherence of the
reflected wave can be met easily. For a nucleon of 20 MeV
energy we find that a 30 MeV well satisfies the condition
if it exceeds 3 Fermi in radius. A nucleus like
$O^{16}$ may very well serve for that purpose. 
The kinetic energy of 20 MeV is above the Coulomb barrier for light
nuclei, hence we do not
expect major distortions in the reflected wave when protons are used.
Conversely, the effect may serve as a method to determine
nuclear well depths (or radii) by merely registering the
dead time between bunches in the reflected wave beam.
The effect may also be tested in atomic collisions at
backward angles.
\vspace{3 pc}

{\bf Acknowledgements}

This work was supported in part by the Department of
Energy under grant DE-FG03-93ER40773 and by the National Science Foundation
under grant PHY-9413872.

\end{document}